# A Spatial-Temporal Spreading Communication Technique and its Applications

Mingyong Zhou
School of Computer science and Communication Engineering,
GXUST
Liuzhou, China
Zed6641@hotmail.com

*Abstract*—In this paper, we outline a communication approach based on spatial-temporal spreading technique. We first start by investigating the CDMA approaches which is widely applied in industry. Secondly we introduces a less often applied technique that is based on time spreading method. While now days most communication standards are based on CDMA such as TD-SCDMA, CDMA2000 etc, time spreading techniques are introduced in this paper to demonstrate the advantages of techniques based on time spreading method. Time spreading technique does demonstrate certain merits in certain scenarios such as impulse noise disturbance. Throughout this paper, we applies Chrestenson (CH) functions that are a set of complete and orthogonal multi-value functions that generalizes the best known Walsh functions. We even demonstrate that certain CH sequence may be used for future PN sequence to avoid Multiple Access Interference(MAI ) problem and synchronization problem in current CDMA systems. Lastly through several simulations we show the advantages of both frequency spreading CDMA and time spreading techniques from various aspects. While CDMA demonstrates certain merits under noise scenarios by spreading the spectrum of signals and thus enhances the channel capacities by Shannon's theorems in reference [1], in this paper we also show that under certain scenarios such as impulse noise environments, time spreading technique demonstrates similar merits and can also enhance the channel capacity under Shannon 's theorems in reference [1], given that signal's frequency bands remain the same and given an appropriate choice of temporal spreading technique . We thus show that with appropriate choice of temporal spreading technique, the bandwidth of the temporal spread signal could be enhanced .

*Key words:* Chrestenson function; DCHT;Spread communication; CDMA;TDMA;

I. INTRODUCTION

Various coding methods are proposed in recent years including Turbo coding whose channel capacity reaches the upper limit that is proved by Claude E. Shannon in 1948. An obvious observation from [1] is that as long as we can spread the transmit signal's bandwidth and at the same time increase the average transmitter power, the communication channel capacity can be increased. Based on this rationale, we investigate in this paper various temporal spreading techniques that could enhance the bandwidth of transmitter signals. Temporal spreading technique is known for its robustness to impulsive noise. But its other advantages were not investigated . In this paper we show that other than robustness to impulsive noise, temporal spreading can increase the bandwidth of transmitted signals by using Chrestenson functions , and thus the channel capacity could be enhanced through operations of the Chrestenson functions. This argument is line with the well-known Shannon's theorems in [1]. In addition, the temporal spreading sequence could be manipulated in a way that encipher is possible.

II, .Chrestenson function and Chrestenson transform

Chrestenson function proposed by H.F. Chrestenson in 1955 is an extension and generalization of Walsh transform. It is defined as follows as outlined in [2].

Definition 1:(p-adic multiplication) for any integer p≥2, for any x∈[0 , +∞ ) , t∈[0 , +∞ ) , their p-adic integers can be represented as：

$$x = (x_{-s}...x_{-2}x_{-1}x_0 \bullet x_1 x_2 ... x_{r+1})$$

$$t = (t_{-r}...t_{-2}t_{-1}t_0 \bullet t_1 t_2 ... t_{s+1})$$

p-adic multiplication of t and x is defined as

$$t \otimes_p x = \sum_{k=-s}^{r+1} t_{1-k} x_k \quad (1)$$

where $\sum_{k=-s}^{r+1}$ denotes mod p sum.

Definition 2: (Chrestenson forward transform)

$$C(k,\omega) \triangleq \exp\left\{-\frac{2\pi j}{p}\omega \otimes_p k\right\},$$

where $k = 0,1,2\ldots \infty$ (2)

is defined as Chrestenson function. For a series $x(n), (n=0,1,\infty)$, its Chrestenson forward transform is defined as

$$X(\omega) \triangleq \sum_{n=0}^{\infty} x(n)\exp\left\{-\frac{2\pi j}{p}\omega \otimes_p n\right\},$$

where $\omega \in [0,1)$ (3)

Where p is an integer number larger or equal to 2. Note that a prime number is required in most references, and in fact in [1], it is shown that p can be any integers larger than 2. $\otimes_p$ denotes a p-adic multiplication defined in Definition 1.

**Theorem 1**: Similar to Fourier theory, it can be verified that the Chrestenson inverse transform is as follows[2]

$$x(n) = \int_0^1 X(\omega)\exp\left\{\frac{2\pi j}{p}\omega \otimes_p n\right\}d\omega \quad (4)$$

The proof is obvious by using the property of basis function and it can be referred in [2]

We have shown in [3] that Discrete Chrestenson Transform (DCHT) is derived as the following pair of equations, where p is any positive integer number.

$$x(n) = 1/p \sum_{k=0}^{p-1} X(k/p)\exp\left\{\frac{2\pi j}{p}(k/p \otimes_p n)\right\}, \quad (5)$$

$$X(n) = 1/p \sum_{k=0}^{p-1} x(k/p)\exp\left\{-\frac{2\pi j}{p}(k/p \otimes_p n)\right\}, \quad (6)$$

## III, TEMPORAL SPREADING USING CHRESTENSON FUNCTION

In Equation (2), Chestenson function can be similar to Fouier's and when p=2 it is Walsh function. In reference [3], we show the relationship between Chrestenson function and Fourier's when p is larger than 2. It is interesting to note that both from theory and simulations that when sample numbers are the same as p, and p is the exponentials of 2's, the Discrete Chrestenson function (DCHT) is equivalent to DFT [3].

The principle of temporal spreading is based on equation(2). For a given x(n) where n=0,1,2..N, we select a given $\omega_1$ and calculate a series of $x(n)C(k,\omega_1)$, where k=0,1,,,K are the temporal spreading points with total number of K for an given x(n) (n=0,1,...N). Therefore the temporal spreading total number is KN if we temporally spread all x samples evenly in the above way. We have the following theorem2 for the above temporal spreading technique.

**Theorem 2**: With the above temporal spreading technique described, if the following condition A) are satisfied:

A) sample numbers and p's are the exponential of 2's.

The bandwidth of temporally spread signal is enhanced to $\omega_{max} + \omega_1$, where $\omega_{max}$ is the maximum angular frequency of original signal x, and $\omega_1$ is the angular frequency in Chrestenson function denoted by Equation(2).

**The proof** is straightforward. First under condition A), the DCHT is equivalent to DFT in reference[1]. When we perform the temporal spreading $x(n)C(k,\omega_1)$, where k=0,1,,,K for each x(n), n=0,1...N. Without loss of generality, suppose x(n)=Acos($\omega_{max} t + \phi$), then for $x(n)C(k,\omega_1)$ its angular frequency is between ($\omega_{max} - \omega_1, \omega_{max} + \omega_1$). This is similar to a RF modulation process in which a speech signal is modulated into a much higher frequency in order to transmit in the air.

Actually condition A) can be relaxed for any given positive integer p when p is 2 or larger than 2 and any sample numbers, but the proof needs more calculation and is omitted here.

We illustrate the principles of theorem 2 in figure 1 in which original cosine signal x(t)'s samples are spread in time to another higher frequency signal.

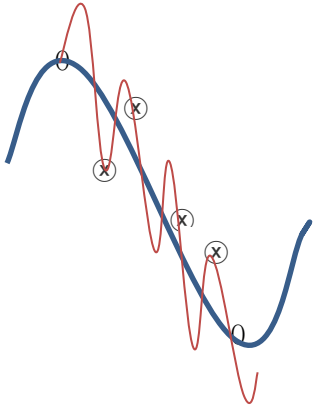

○ denotes lower frequency original signal's samples

⊗ denotes temporal spread samples with higher frequency

Figure1: bandwidth is increased :original signal x(t) with lower frequency and its temporal spread signal with higher frequency.

Following the theorems in reference [1], we conclude that the channel capacity will be enhanced with the upper and lower boundary after temporal spreading as described . More importantly, each samples can be recovered by exploiting the property of Chrestenson function.

IV. Spatial and temporal spreading techniques

In fact we can spread the signal both temporally and spatially in frequency in different sequence  so that the a secure communication can be designed. A straightforward approach is temporally spread the original signal in this paper and then spatially spread the temporal spread signal using PN seqence or CH sequence proposed in our recent  report, or vice versa . But using the temporal spread at a later stage obviously will enhance the robustness to impulsive noise disturbance in transmission channels. It is expected that this temporal-spatial spread technique has more merits in terms of security as well as channel capacity. An important note is that temporal spreading can actually enhance the channel capacity as shown in this paper other than its robustness to impulsive noise disturbance.

V. Simulations

We present simulations where an impulsive noise disturbance exists in the transmission channels as well as the bandwidth changes after the temporal spread.

Figure 2 denotes the error signal after a random impulsive noise is added with noise amplitude being between 10% and 100% of temporal spread signal's amplitude. Every 10 samples a sequence of impulsive noise is added.

Figure 3 denotes the waveforms of one of recovered signal and its original signal. Note that they are almost in line with each other in simulations.

Figure 4 is the power spectrum of one of the original signals, note that it decays at higher frequencies where 128 samples are used for calculation..

Figure 5 is the power spectrum of temporal spread signal using Chrestenson function where p=8, 128 samples are used. It is worthy to note that the temporal spread signal bandwidth is enhanced to cover almost all frequency bandwidth, a property that is close to white noise and is what we expect.

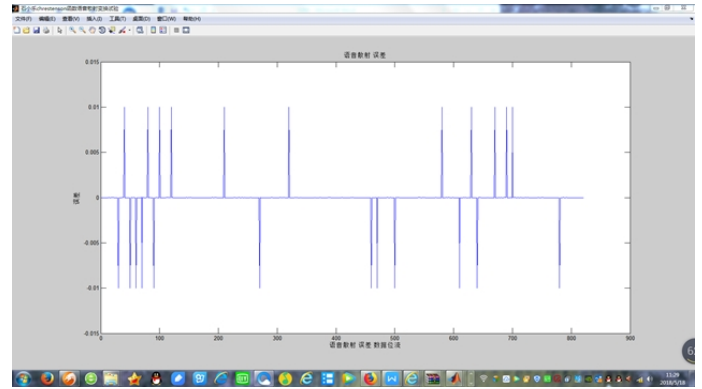

Figure 2: the error signal of recovered signal as compared with original signal

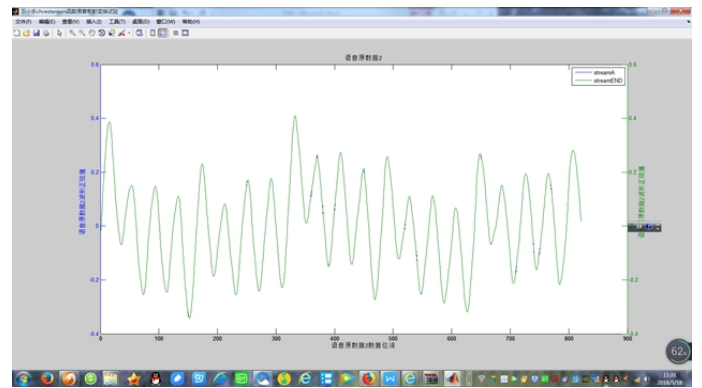

Figure 3: Waveform of one of the recovered signal vs its original signal

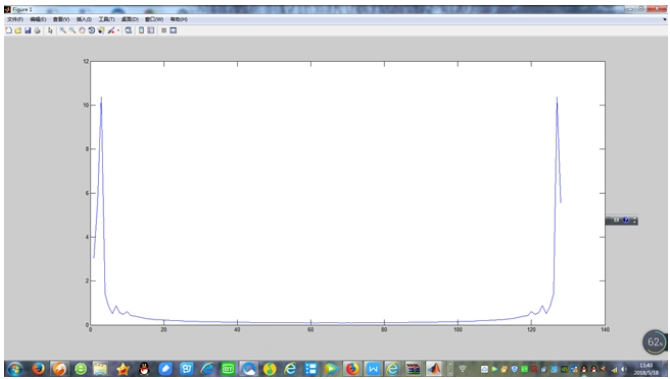

Figure 4: Power Spectrum of one of original signals

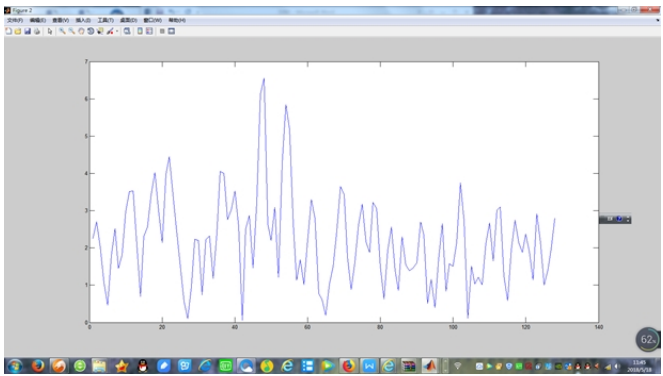

Figure 5: Power Spectrum of temporal spread signal with wider bandwidth

VI. Conclusion and Further considerations

In this paper, we propose a temporal-spatial spread communication method by using Chrestenson functions. We first introduce the properties of Chrestenson functions and Chrestenson transform. Then by making use of its multiple value property we illustrate a spread communication method in which the data streams are first temporal spread, mixed and then transmitted. At the receiver side, the mixed streams are decoded and recovered. Finally we show a voice simulation on MATLAB to show the effectiveness of this method. We show in theory and simulation that the transmitted signal bandwidth is enhanced close to that of white Gaussian noise, and thus the channel capacity is maximized according to the Shannon's theory. In fact, Chrestenson function is one of many approaches that can be used to enhance the bandwidth of temporal spreading signals among many others such as binary Walsh function ,Continuous Fourier function for example. Another point to note is that in our simulations the channel coding methods are not taken into considerations. If proper channel coding method such as Turbo coding ,Hamming coding is used, the errors in Figure 2 will be largely reduced close to zero or to zero. In [1], an example is demonstrated in which transmission rate is equal to the channel's Shannon capacity ,a method due to R. Hamming. In this sense, our paper proposes a method to enhance the signal to maximum bandwidth close to that of white Gaussian noise first, thus enhance the capacity of channel. This result is line with the observation in [1] that to reach the highest transmission rate close to the channel capacity, transmitted signal must be close to white Gaussian noise. Then we can employ existing channel coding methods such as a method by R. Hamming to code the signal at a transmission rate equal to the channel's Shannon capacity.

**Acknowledgement**

The author would like to thank Xiaole Shi for providing his help during simulations.

**References**

1, Claude E. Shannon, "A Mathematical Theory of Communication", The Bell System Technical Journal, Vol.27,pp.379-423, 623-656, July ,October, 1948.

2, H.F. Chrestenson,"A Class of generalized Walsh transform ", Pacific Journal of Mathematics, Vol.5, PP.17-31, 1955.

3,Mingyong Zhou et al, " *Chrestenson transform and its relations with Fourier transform*, The Third International Conference on Robot, Vision and Signal Processing, November 18-20, 2015 Kaohsiung, Taiwan. ( IEEE Computer Society Publishing )